# Metallic Hydrogen: Experiments on Metastability

W. Ferreira[1], M. Møller[1], K. Linsuain[1], J. Song[1*], A. Salamat[2], R. Dias[3], and I.F. Silvera[1]
[1]Lyman Laboratory of Physics, Harvard University, [2]Dept. of Physics and Astronomy, University of Las Vegas, Nevada, and [3]Dept. of Physics and Astronomy, Univ. of Rochester

Molecular hydrogen was pressurized in a diamond anvil cell at temperatures between 5 and 83 K. At a sufficiently high pressure, estimated to be between 477 to 491 GPa, hydrogen became metallic, determined by its reflectance in the near infrared and fit to a Drude free-electron model. We then studied the predicted metastability of metallic hydrogen. At a temperature of 5 K the load on the metallic hydrogen was stepwise reduced until the pressure was zero. While turning the load or pressure down, the sample evidently transformed to the molecular phase and escaped; the sample hole closed. We estimate this pressure to be 113 to 84 GPa. Metallic hydrogen was not observed to be metastable at zero pressure.

In this paper we study metallic hydrogen (MH) at high pressures and low temperatures in a diamond anvil cell (DAC). There are two objectives: 1. Reproduce MH in the laboratory and 2. Study the possibility that MH is metastable, or answer the question "Will MH remain metallic when the ultrahigh pressure needed to produce it is lifted, or will it convert back to molecular hydrogen?"

Metallic hydrogen can be made along two pathways: a low-temperature pathway, Pathway I, at very high-pressure and a high-temperature pathway at intermediate pressures, Pathway II (see the phase diagram in the SI; Fig. S1 from Ref. [1]). Pathway II is a first-order phase transition to liquid atomic metallic hydrogen. The phase line for the latter was observed by Zaghoo, Salamat, and Silvera [2] at pressures of order 100-200 GPa and temperatures 1000-2000 K. Pathway II is not of relevance for this letter.

Pathway I, is the focus of this paper. In 1935 Wigner and Huntington [3] predicted that at a sufficiently high density or pressure, solid molecular hydrogen would transition to atomic metallic hydrogen by molecular dissociation. This insulator to metal transition was first observed in the laboratory by Dias and Silvera [4] at a pressure reported as $495 \pm 13$ GPa at temperatures of ~5 $K$ and 83 $K$. The pressure was increased in steps, while observing the sample to transition from a transparent insulator, to opaque black, and finally to shiny MH, surrounded by the less reflective rhenium (Re) gasket, shown in Fig. S2. The reflectance was measured as a function of wavelength and fit to a Drude free-electron model of a metal that yields the plasma frequency or the electron density. It was found to be a metal with one electron per proton or atomic metallic hydrogen.

This first observation of the Wigner-Huntington transition to atomic MH was met with comments and objections, but all could be answered [5]. We agreed with two of the comments: that the experiment on MH should be repeated, and that the Drude model should be fit for more than two wavelengths. We set out to respond in two ways: measure the reflectance of MH (the subject of this paper) and measure the electrical conductivity of MH. The reflectance was measured at Harvard and the conductivity at the U. of Rochester [6]. In the latter, the transition pressure was measured to be 470-480 GPa, using the shift of the Raman active phonon in the diamond, currently the best method of determining ultra-high pressures in DACs. This is in agreement with the pressure determined by Dias and Silvera. In the present paper we measured the reflectance at 5 wavelengths in the visible/infrared, confirming the metallic behavior of hydrogen. We have had a few other observations in which MH was produced. For technical reasons, nothing new was measured, so the resulting repeated observation of MH was not published, except to show photos of the MH phase [7].

*Current address: Institute of Physics, Chinese Academy of Sciences, Beijing 100190, China.

If MH exists at ambient conditions and is superconducting it would be revolutionary, with applications having a large impact on society [8]. Metastability of MH was first proposed 50 years ago by Brovman, Kagan, and Kholos [9]. They carried out a detailed calculation of the energy and equation of state, including possible structures, using perturbation theory in the low temperature limit. They found that the ground state could be a liquid, or an array of protonic filaments (one dimensional chains), yet a stable metal at low pressure. They considered the effect of zero-point motion (ZPM) and realized that metallic deuterium with a much smaller ZPM had distinct advantages for metastability, compared to hydrogen. They also considered that a short lifetime of the metastable phase might limit its usefulness. On a fundamental level, if MH is a low temperature liquid it might have novel quantum states, with both superfluidity and superconductivity [10].

A recent discussion of metastability with an extensive list of references is provided by Tenney, Sharky, and McMahon [11]. They carried out a first principles analysis of this problem using density functional theory. There are two considerations for metastability: One must first find the metastability pressure $P_m$ in the low temperature limit. If the sample is metastable at ambient pressure, then the temperature should be increased until thermal dynamics causes a transition to molecular hydrogen at the temperature $T_m$. For example, diamond is a metastable phase of carbon. $P_m$ is zero atm.; however, if heated to a temperature $T_m \approx 1400\ K$, diamond will convert to the low energy phase, graphite. Tenney et al find $P_m \approx 200\ GPa$ for MH. Burmistrov and Dubovskii [12] predict stability down to 10-20 GPa, while Ackland [13] finds MH to be unstable at 0 atm. Clearly, an experiment should be carried out to determine $P_m$ and $T_m$. In the present experiment, we considered T ≈ 5 K to be in the low temperature limit.

In order to observe metastability, our plan was to load hydrogen in a DAC, make MH and confirm this by measuring reflectance at liquid helium and nitrogen temperatures. Then, at a sample temperature of $5K$, we would slowly release the load while observing the sample. If it remains in the gasket hole at effectively zero pressure, then it is metastable; if it disappears it is not.

With our DACs we can continuously raise the pressure by turning a screw to increase the load on the diamonds and there is a linear relationship between load and pressure. When the load is lowered (turning the screw back) the pressure is reduced. In several runs on molecular hydrogen in the past, we studied the sample with decreasing load. At a certain point, we observed that molecular hydrogen escaped and the hole, while still under some load, closed up. For the present experiment we decided not to measure the pressure by observing the Raman active diamond phonon, as this could lead to catastrophic failure of the diamonds, with no possibility of observing the sample and metastability. The transition pressure of MH has already been measured a few times, so it need not be repeated. In 1998 Narayana et al [14] studied hydrogen to 342 GPa using Raman scattering of the hydrogen vibron. Fifteen sets of diamonds broke every time the stressed diamonds were illuminated with laser light used for the Raman scattering (A. Ruoff, Pvt. Comm. to I. Silvera).

We cryogenically loaded 0.9999 pure hydrogen in a DAC. The diamonds were conic type IIas, with 30 μm culet flats and a 9 degree bevel with a diameter of 300 μm. Much of our results are from visual observation of the sample. We used two microscopes in this study. The Leitz microscope was used for preparation of DACs at room temperature, with a magnification of a few hundred. For viewing the sample in the cryostat (which has 3 $CaF_2$ windows intervening) we used a Wild Macroscope, with a high quality long-working distance objective (120 mm) with magnification of 30 to 40.

In our DACs the load (the force on the diamonds) is increased by rotating the drive screw on the DAC with a long stainless-steel tube that extends out of the cryostat at room temperature. Our DACs are also provisioned with calibrated strain gauges that indicate the load or force. We have found in many previous runs that the measured pressure is proportional to the rotation of the screw or the load. We indicate the number of turns in 8ths of a turn; we also record the load from the strain gauge. In the present experiment many of our observations are recorded with photos indexed with the number of turns.

The pre-indented gasket was made of rhenium, with a 19 $\mu m$ diameter sample hole and thickness of 10 $\mu m$ under the culet. The gasket was gold coated before indenting to minimize stress induced damage to the diamonds. The hole was made by electric discharge machining (EDM) which leaves a ring of black slag (melted and redeposited metal) around the edge of the hole (see Fig. 1a). Our usual procedure is to use high-power ultrasonics on the gasket submerged in water mixed with micron sized alumina particles; this removes the gold, but not the slag. The slag is then knocked off with a sharpened toothpick. The diamonds and gasket are then coated with a 50 $nm$ layer of alumina; this protects the diamonds from hydrogen diffusion and embrittlement, while the Re gasket is protected from interacting with hydrogen to form rhenium hydride.

This last step to remove the slag was inadvertently skipped; some of the slag was then crushed between the two diamonds but the slag also spread out over the rhenium leaving a possibly rough surface. Consequently, the slag could cause leakage of the hydrogen until fully flattened during cryogenic loading. When the sample was loaded, the initial diameter was about 10.8 $\mu m$ rather than closer to 19 $\mu m$, resulting in a smaller than desired initial sample diameter. Moreover, the black slag smeared out on the Re so that when the sample was viewed through a microscope, rather than seeing a transparent hole surrounded by reflective Re, the hole was surrounded by a black smear of slag (see Fig. 1b). The sample shape is sometimes more elliptical than round. We measure the area and then provide an effective diameter, i.e., the diameter of a circle with the same area as the sample.

If one indents a gasket there is an almost linear pressure gradient from the culet center to the edge of the culet where the pressure drops to very low values. Thus, if a sample like hydrogen is loaded, as pressure is increased the hole not only gets smaller but can migrate from the center of the culet to the edge. That is what happened in this case (see Fig. 1c). In this figure the load had been increased so that the small sample ($diam.\sim 3$ μm) is at the top edge, indicated by an arrow and surrounded by Re. Rhenium is also exposed at the lower part of the figure.

Our DACs have diamonds mounted on cylinders that fit snugly in a cylindrical hole (see Ref. [15] for the design). One cylinder is fixed and the other translates to increase the load when the screw is turned. In our first observation of MH [4], when the screw was rotated the pressure increased. Subsequently, in runs with this DAC, at high pressure, when the screw was turned the load would increase, but sometimes the pressure would not change until after a few turns, increasing the force, and then the pressure jumped up, while the load, measured with the strain gauge, partially fell back. The source of this sticking problem is under investigation. At about 240 GPa, with turning of the screw the pressure increased to the $H_2$-PRE pressure region, i.e. above about 360 GPa. With further turning the sample turned black, and finally began to shine, that is, it became metallic (see Figs. 1c and S3).

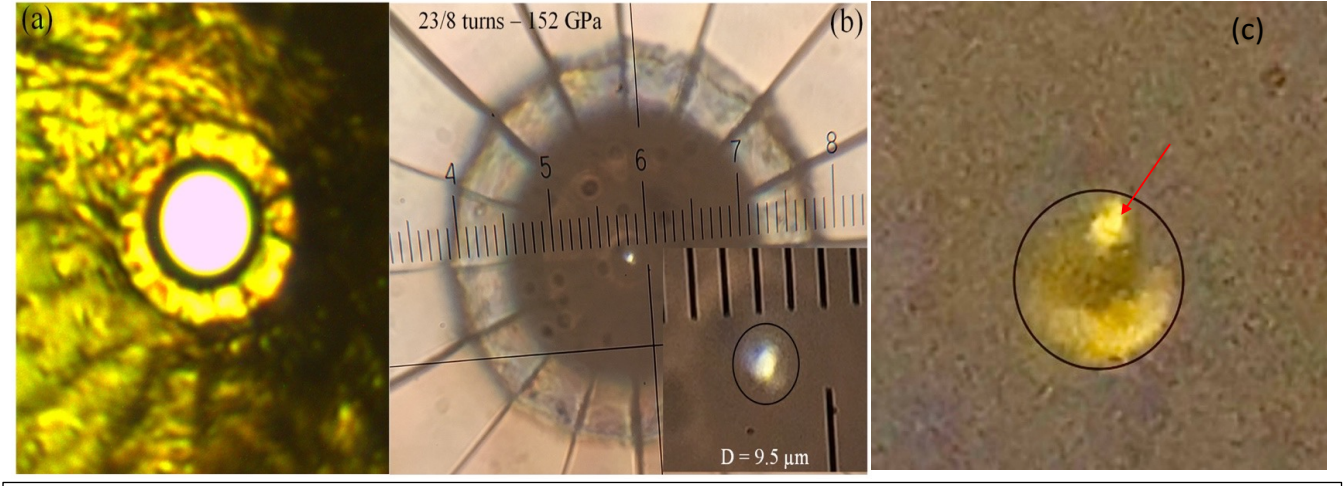

Figure 1. a) The pre-indented gasket viewed with the Leitz microscope, back and front-lit to show the EDM'd hole; note the black slag at the edge of the hole. b) The sample under the diamond viewed with the Macroscope. This high-resolution image can be zoomed in (inset) to measure the sample hole size. We have superimposed a circle around the 30 µm diameter culet. Note that the sample hole is surrounded by black slag, rather than reflective Re. c) Photo of MH at very high pressure (46/8 turns).

To measure reflectance, one measures a reflected intensity and normalizes to the incident intensity. The incident intensity is challenging to measure in a DAC as the sample cannot be removed. As shown in a recent paper on the reflectance of rhenium [16], great care must be taken to properly measure reflectance in a DAC. To measure the reflectance $R_{ds}$ of the light off the sample, $s$, pressed against the diamond, $d$, one needs to know the intensity $I_0$ of the light incident on that surface. We measure the light $I_d$ reflected off the diamond table and the light $I_{ds}$ reflected off the MH sample, pressed against the diamond. The incident intensity $I_0$ propagates into the diamond, reduced by $R_d$, the reflectance from the diamond and then reflects off the sample/diamond surface. This reflected light is again reduced by $R_d$ as it propagates out of the table to a detector that measures the intensity. It is easily shown that

$$R_{ds} = \frac{R_d}{(1-R_d)^2} \frac{I_{ds}}{I_d} e^{2\alpha D}$$

The exponential term is due to absorption or scattering of light out of the beam in the diamond of thickness D and absorption coefficient $\alpha$; the factor of 2 is due to the double pass through the diamond. $R_d = (n_d - 1)^2/(n_d + 1)^2$; here $n_d = 2.41$ is the index of refraction of diamond in the near IR.

To measure $I_d$ and $I_{ds}$ (see the optical layout in Fig. S6), we used 5 narrow-band light emitting diodes (LEDs) at wavelengths 625, 720, 810, 890, and 950 nm, obtained from Digi-Key. These were mounted on a disc and could be rotated into position. The light was collected with a lens and focused on a pinhole; the emerging light was collimated with a lens and directed into the Macroscope to be focused into the DAC. The optical set-up was on a platform that could be adjusted with micrometers to focus into the DAC, mounted on a large optical table. The cryostat itself could be rotated or tilted so that the light was normal to the diamond table. The LEDs had to be carefully aligned to avoid large variations in the measured reflectance. Only light from the 625 nm LED was easily visible to the eye. To align the other LEDs, we placed an IR sensitive camera behind the collimating lens and adjusted the LED until the intensity was symmetric and almost uniform. Full alignment took several hours.

In this analysis, the major unknown was the absorption of the diamond. Vohra [17] has studied the opacity of a few types of diamonds under load. At that time it was thought that the opacity of diamond was due to closure of a band gap, but recent studies at SLAC [18] conclude that the gap does not close and the opacity is due to scattering or interaction with impurities. Nevertheless, we can consider the data of Vohra who plots the optical density (OD) (see Ref. [4], Fig. S5). We extrapolate this data to 500 GPa and then interpolate to find the opacity in the IR (Fig. S5). This indicates that the OD is rather flat in the near IR. However, the values of about 1.5 would give unphysical values or reflectance far greater than 1 in Eq. (1). We believe that our diamonds are of very high quality and would have a lower OD; we used a value of 0.4 for all LEDs, to have a physically acceptable result with reflectance less than 1.

For measurements of the reflected signal, we used a monochrome 1440x1080 pixel CMOS camera with square 3.45 $\mu m$ pixels, model CS165MU, from Thorlabs. It is sensitive in the near IR, to just beyond 1000 nm. The advantage over a color CMOS is that there are no Bayer filters, so the overall quantum efficiency is greater at any detectable wavelength. In monitoring the progress of the hydrogen sample, it is useful to have a color camera, so we took smartphone photos of the sample at all loads.

To determine the uncertainty in reflectance, we misaligned the LEDs then realigned to again measure. This was done twice at liquid nitrogen temperatures to determine an uncertainty. Reflectances were only measured once at helium temperatures, due to the recent liquid helium shortage and limitations on time with a cold cryostat, so that we had enough helium reserve to study metastability at low temperature. We are not sure of the uncertainty of reflectance at helium temperatures as we did not have sufficient time (liquid helium) to carry out our usual alignment; this data was not fit.

The first result is the progression of the sample pressure and size with increasing load. Figure 2 shows the effective diameter of the sample as a function of load or number of turns. The sample of MH was very small, about 3 $\mu m$ in diameter. With increasing load we could also measure the pressure in some cases. This is shown in Fig. S4 where the sample progresses in pressure; at 23/8 turns it enters phase III (see Fig. S1). At 29/8 turns the pressure jumped up and the vibron absorption disappeared. The sample entered into phase $H_2$-PRE. In Fig. S4c we show the fall off of the integrated intensity, similar to observations made in Ref. [1], Fig. 2c; this corresponds to the closing of the electronic gap.

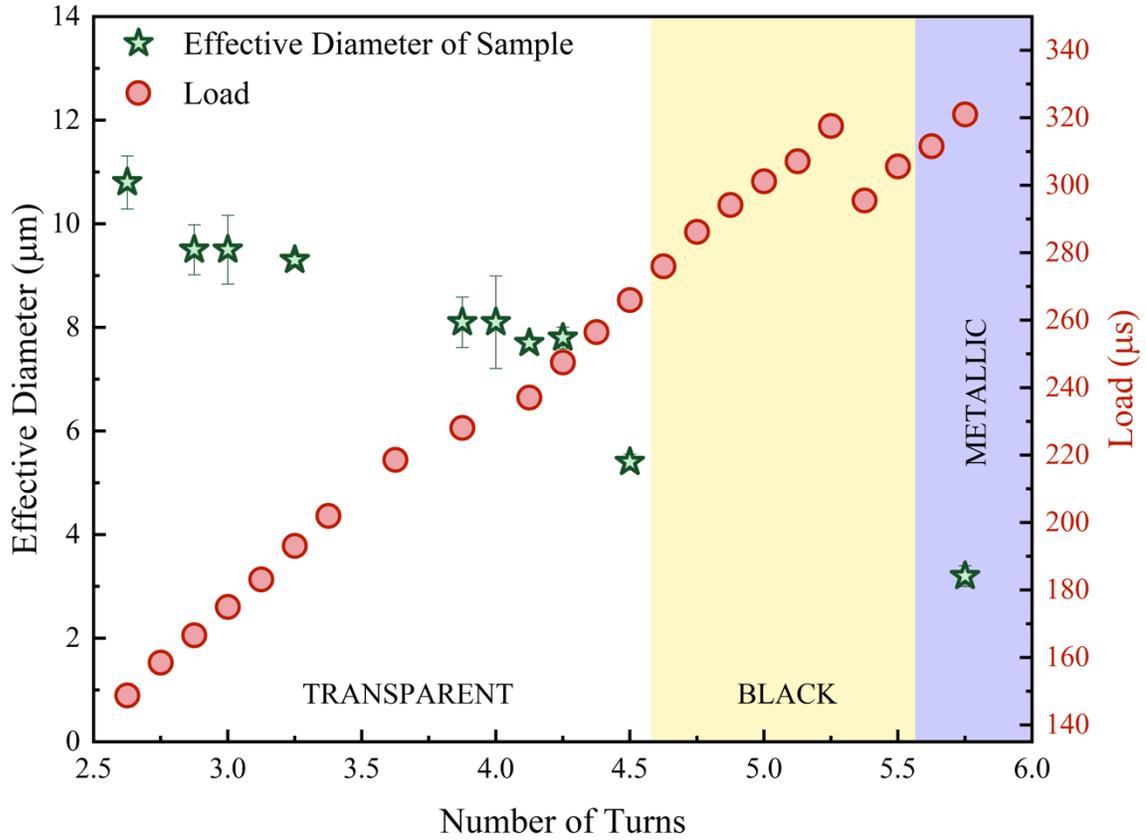

Fig. 2. The effective diameter of the hydrogen sample as a function of number of turns of the screw, or load. The screw was turned up by eighths of a full turn, for a total of almost 6 full turns. The sample progressed from transparent to black to metallic. A kink in the load (red points) usually corresponds to a jump in pressure.

The measured reflectance at liquid nitrogen and liquid helium temperatures is shown in Fig. 3. The basic trend is for the reflectance to increase as the photon energy decreases. The only slight deviation is for the 1.3 eV point (950 $nm$~1mm). The MH sample had migrated to the edge of the culet and was surrounded by Re (note the black slag). The elliptically shaped sample had a small dimension of about 2.8 $\mu m$ and the reflectance of MH could be contaminated with reflectance from the Re. From the fit, the plasma frequency is consistent with atomic metallic hydrogen [4] while the scattering time $\tau$ is much longer than found by Dias and Silvera, perhaps indicating fewer scattering centers in the sample. The data confirms that MH has been reproduced.

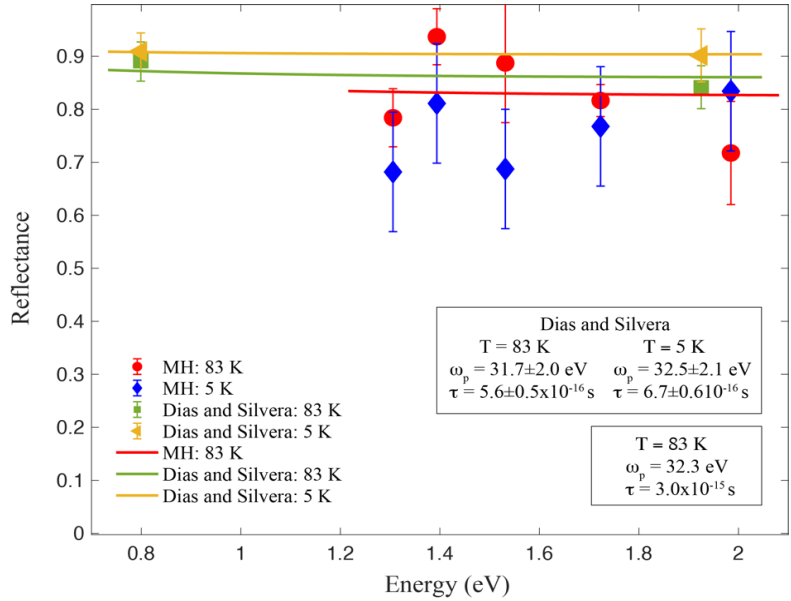

Figure 3. Reflectance of MH as a function of photon energy. We compare to that measured by Dias and Silvera. Uncertainties are described in the text. Data at helium temperature is shown, but was not optimized, due to helium limitations. Pressure was estimated to be in the range 477 to 491 GPa.

To test for metastability we illuminated the sample from both front and back at 5 K. Both a smart phone camera and the CMOS camera were mounted on the eyepieces of the Macroscope and set to run continuously (smartphone: ~30 Hz frame rate; CMOS ~20 − 35 Hz) as the load was reduced in steps of 1/8 turn. Ideally if metastable, we would see the MH sample at zero pressure and then raise the temperature until it converted to molecular hydrogen. We zero'd the strain gauge meter (set 46/8 turns to zero, Fig. 2,) and measured the release of strain as the load was reduced. In Fig. 4, at 16/8 turns release, we could still see the MH, but with more turns it disappeared (between 23/8 and 25/8 turns release), and the hole closed, as we saw no transmitted light. The shiny lower part in the figure is Re. The fact that the hole closed indicates that there was still a load on the sample when it escaped. We estimated the pressure by comparing the release turns to a linearized and extrapolated scale of pressure when increasing the pressure. This extrapolation yielded 477 to 491 GPa for the pressure when MH appeared, in agreement with other measurements of the metallization pressure. From this scale, we estimate the metastability pressure to be between 113 to 84 GPa, but this could have a much larger uncertainty as pressure was inferred, not measured. We continued to release the load to make sure that the pressure was zero and diamonds were no longer compressing the gasket. At this time we can claim with no uncertainty that MH is not metastable at zero pressure.

In this experiment we have accomplished our two objectives. We have reproduced metallic hydrogen in the laboratory, and we have shown that MH is not metastable when the pressure is reduced to zero. This is not the end of the quest for metastability at zero pressure. In the future we plan to metallize deuterium and study its possible metastability.

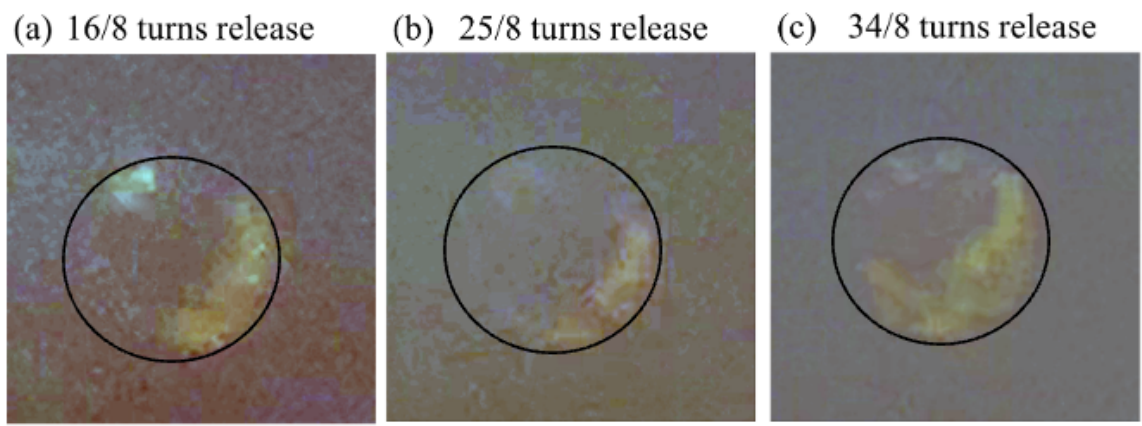

Figure 4. Observation of the sample while the load is released, using a smartphone camera. a) We could still observe the MH, near the upper edge of the culet. b) Sample escaped and hole closed while under load. c) Load or pressure on the gasket is zero. A circle is drawn around the culet flat for ease of viewing.

This research was supported by the SSAA of the DoE with grants DE-NA0003917 and DE-NA0004087 and BES Grant DE-SCF0020303.  Support from the NSF, grant DMR-190594 and DMR-1809649**.**  This work was performed in part at the Harvard University Center for Nanoscale Systems (CNS); a member of the National Nanotechnology Coordinated Infrastructure Network (NNCI), which is supported by the National Science Foundation under NSF award no. ECCS-2025158. KL received partial support from an HCRP (Harvard College Research Program) grant.

# Supplementary Information

# Metallic Hydrogen: Experiments on Metastability

W. Ferreira[1], M. Møller[1], K. Linsuain[1], J. Song[1], A. Salamat[2], R. Dias[3], and I. Silvera[1]
[1]Lyman Laboratory of Physics, Harvard University, [2]Dept. of Physics and Astronomy, University of Las Vegas, Nevada, and [3]Dept. of Physics and Astronomy, Univ. of Rochester

Our supplementary information consists of a collection of figures, shown below.

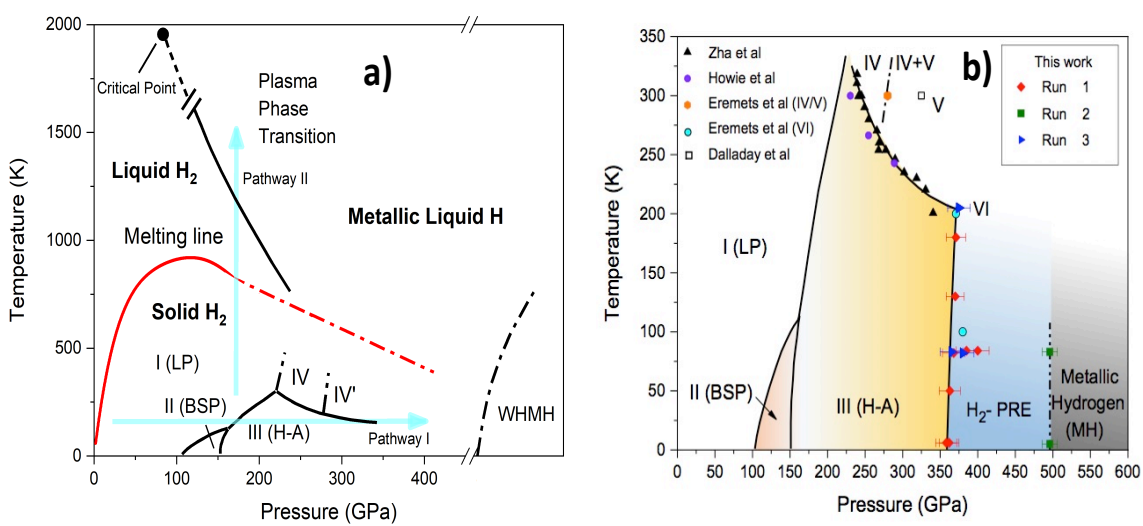

Fig. S1. The pressure/temperature phase diagram of hydrogen from Ref. [2]. a) The high temperature phase diagram showing Pathways I and II to atomic MH. b) An expanded view of Pathway I, showing several phases within the molecular solid, encountered with increasing pressure.

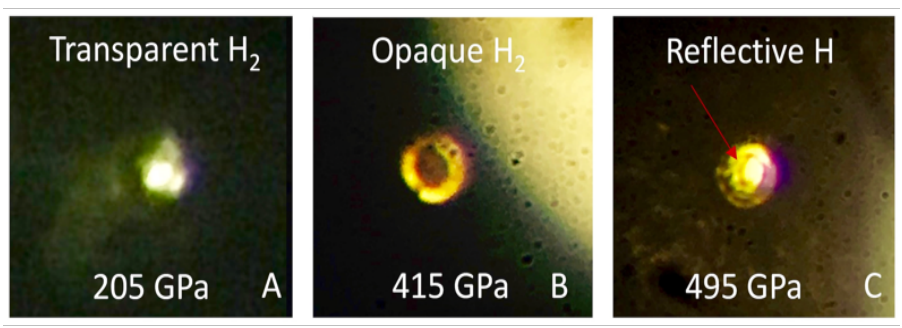

Fig. S2. A figure from Ref. [1] showing hydrogen at various pressures, including MH in C. The sample is back lit in all figures, but B and C are non-transmitting. A is only back lit, while B and C are also front lit.

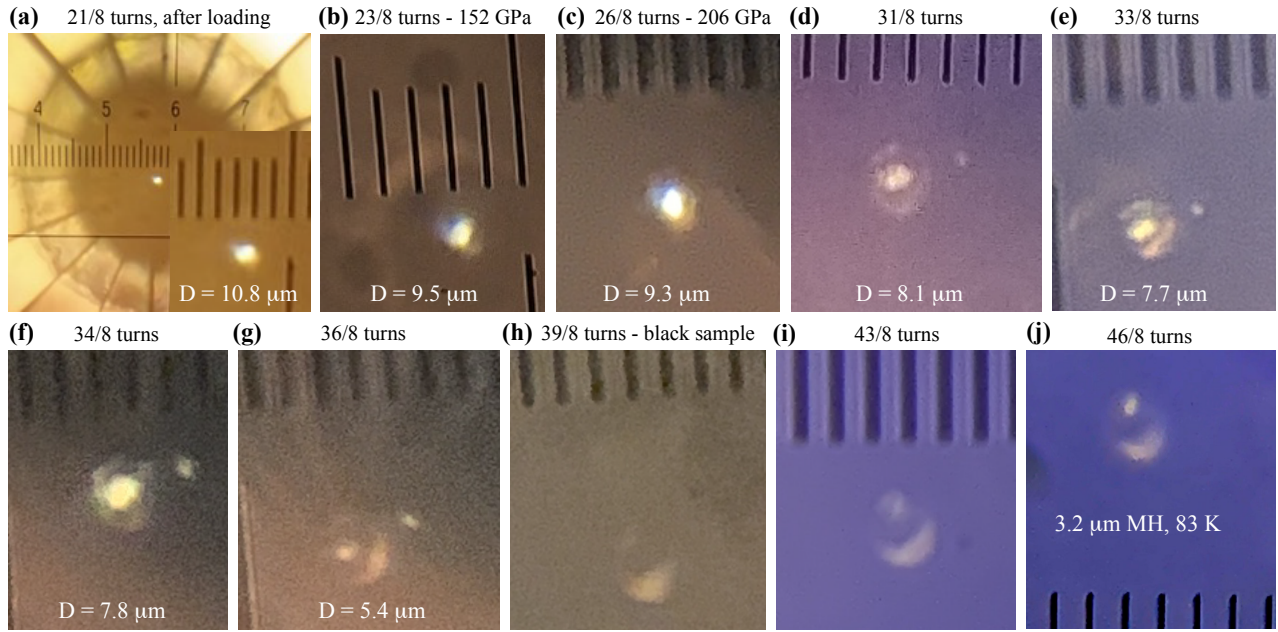

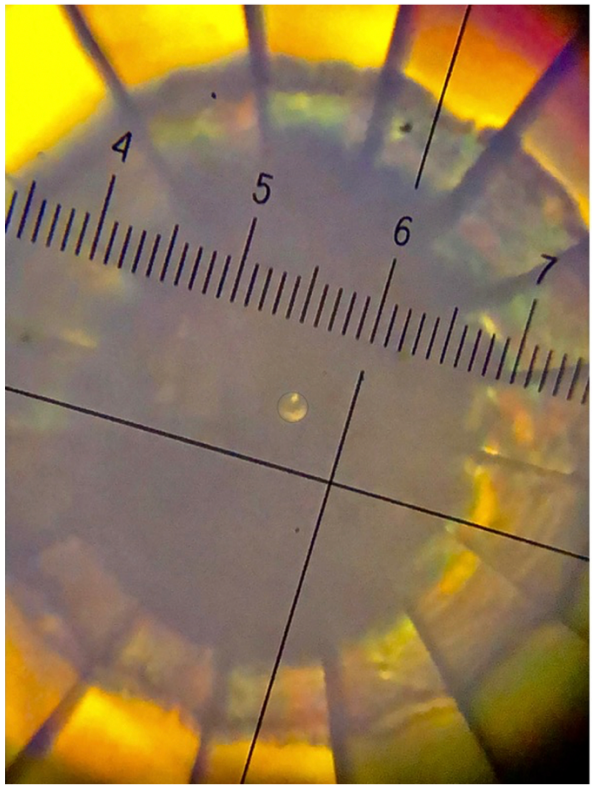 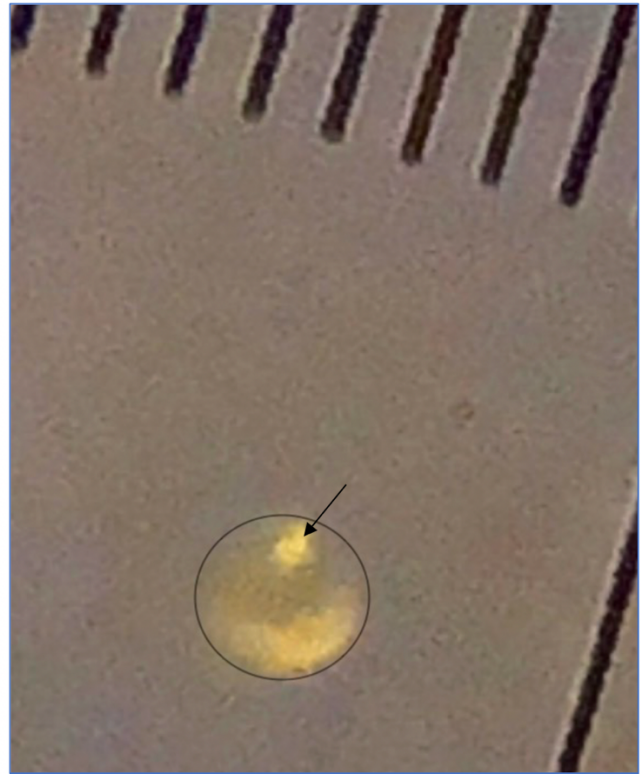

Fig. S3. Photos of the sample with increasing load (turns), corresponding to regions shown in Fig. 2. The upper figure: This shows the progression of the sample as the load was increased. For lower number of turns the sample is surrounded by black slag. With increasing load (e), both the sample and Re are observed. In h) the sample is black and blends in with the black slag. Finally, in j) the sample becomes metallic and shines; in the lower region of the culet the shiny surface is Re.
The lower figure: This shows another picture of the MH, blown up on the right. A circle is drawn around the diamond culet and an arrow points to the MH, surrounded by Re.

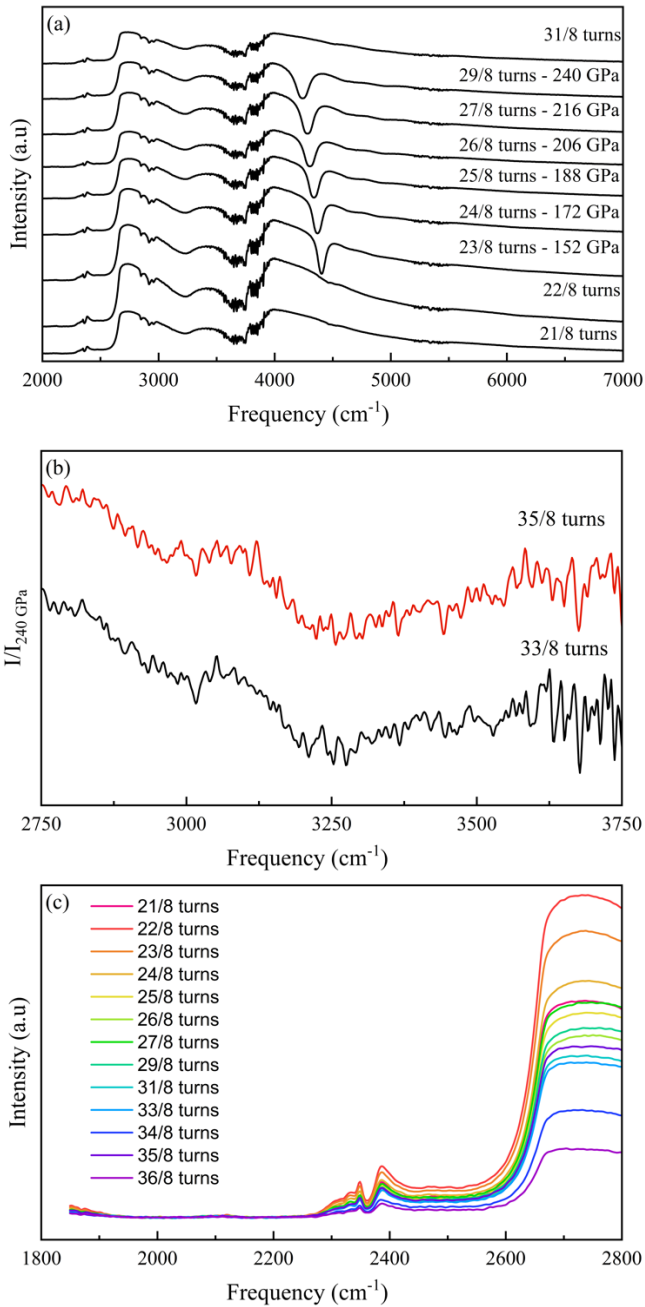

Figure S4. Infrared observations of pressure in the sample. a) This shows the vibron spectrum up to 240 GPa in phase III (see Fig. S1). b) The pressure jumps and the sample enters the $H_2$-PRE phase. c) This shows the integrated intensity in the IR demonstrating the pressure is increasing as the load increases. Comparable results were observed in Ref. [2], Fig. 2.

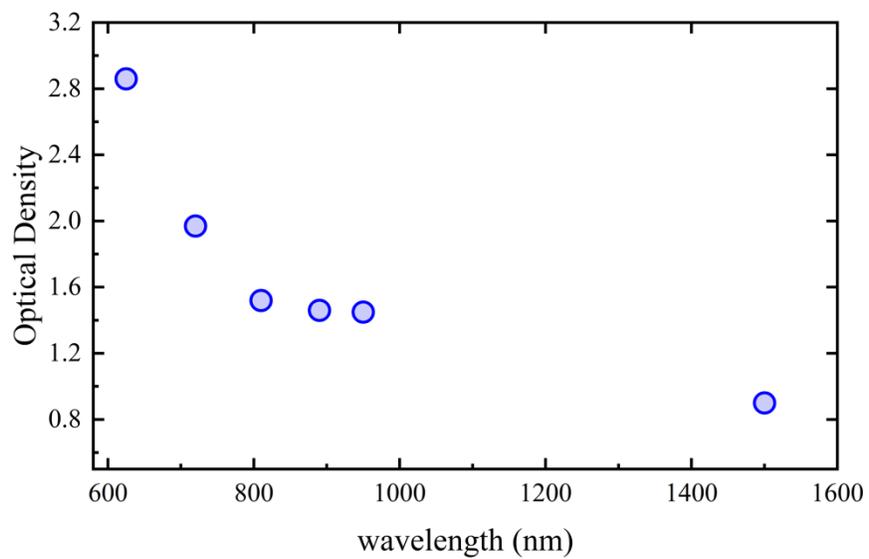

Fig. S5. The optical density extrapolated from Vohra's paper, described in the text.

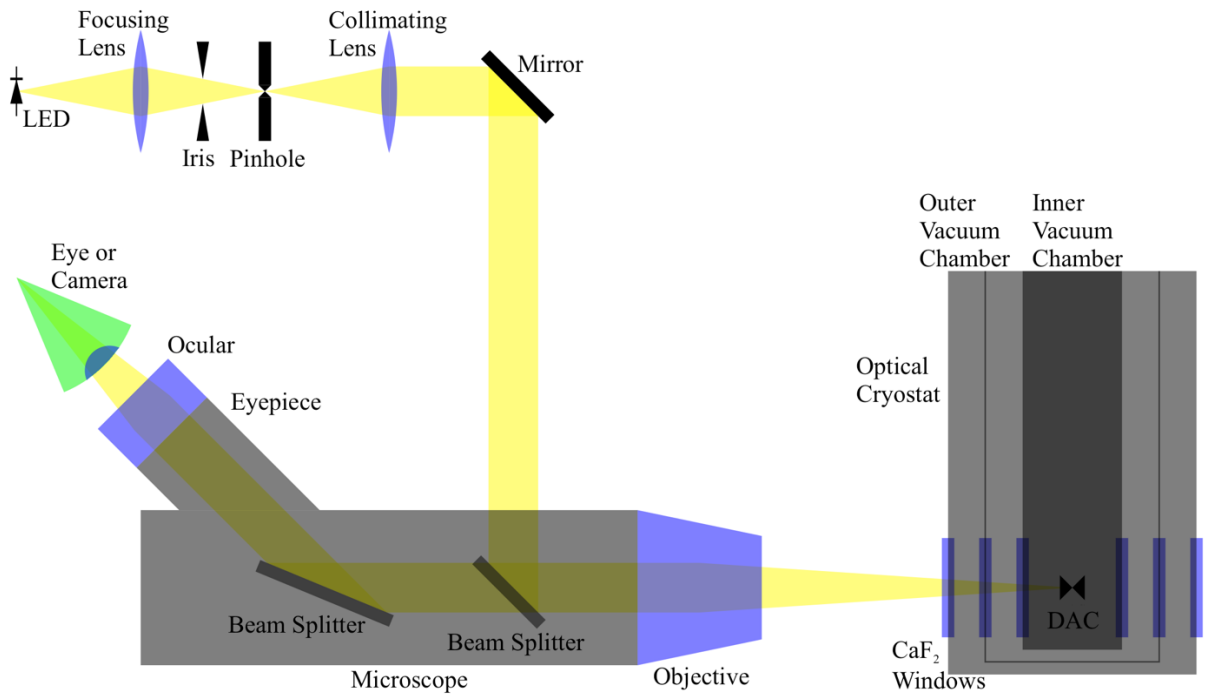

Fig. S6. The optical layout used for measuring the reflectance. Cameras could be mounted on the eyepieces of the microscope, with or without the ocular.